\documentclass[aps,prl,twocolumn,showpacs,superscriptaddress]{revtex4-1}

\usepackage{bm}
\usepackage{amsmath}
\usepackage{amsfonts}
\usepackage{amssymb}
\usepackage[pdftex,
	pdfauthor={Simon Ward},
	pdftitle={Spin Ladder PRL},
	colorlinks
	]{hyperref}
\usepackage{graphicx}

\begin{document}

\title{Bound states and field-polarized Haldane modes in a quantum spin ladder} 

\author{S. Ward}
\affiliation{Laboratory for Neutron Scattering and Imaging, Paul Scherrer 
Institut, CH-5232 Villigen PSI, Switzerland}
\affiliation{London Centre for Nanotechnology and Department of Physics and 
Astronomy, University College London, London WC1E 6BT, United Kingdom}
\affiliation{Department of Quantum Matter Physics, University of Geneva, 
CH-1211 Geneva, Switzerland} 

\author{P. Bouillot}
\affiliation{Department of Medical Imaging and Information Sciences, 
Interventional Neuroradiology Unit, University Hospitals of Geneva, 
CH-1211 Geneva, Switzerland}
\affiliation{Laboratory for Hydraulic Machines, \'{E}cole Polytechnique 
F\'{e}d\'{e}rale de Lausanne, CH-1015 Lausanne, Switzerland} 

\author{C. Kollath}
\affiliation{Department of Quantum Matter Physics, University of Geneva, 
CH-1211 Geneva, Switzerland} 
\affiliation{HISKP, University of Bonn, Nussallee 14-16, 53115 Bonn, Germany}

\author{T. Giamarchi}
\affiliation{Department of Quantum Matter Physics, University of Geneva, 
CH-1211 Geneva, Switzerland} 

\author{K. P. Schmidt}
\affiliation{Theoretische Physik I, Staudtstrasse 7, FAU Erlangen-N\"urnberg, 
91058 Erlangen, Germany}

\author{B. Normand}
\affiliation{Laboratory for Neutron Scattering and Imaging, Paul Scherrer 
Institut, CH-5232 Villigen PSI, Switzerland}

\author{K. W. Kr\"amer}
\affiliation{Department of Chemistry and Biochemistry, University of Bern, 
CH-3012 Bern, Switzerland} 

\author{D. Biner}
\affiliation{Department of Chemistry and Biochemistry, University of Bern, 
CH-3012 Bern, Switzerland} 

\author{R. Bewley}
\affiliation{ISIS Facility, Rutherford Appleton Laboratory, Chilton, Didcot, 
Oxford OX11 0QX, United Kingdom} 

\author{T. Guidi}
\affiliation{ISIS Facility, Rutherford Appleton Laboratory, Chilton, Didcot, 
Oxford OX11 0QX, United Kingdom} 

\author{M. Boehm}
\affiliation{Institut Laue Langevin, 6 rue Jules Horowitz BP156, 38024 
Grenoble CEDEX 9, France} 

\author{D. F. McMorrow}
\affiliation{London Centre for Nanotechnology and Department of Physics and 
Astronomy, University College London, London WC1E 6BT, United Kingdom}

\author{Ch. R\"uegg}
\affiliation{Laboratory for Neutron Scattering and Imaging, Paul Scherrer 
Institut, CH-5232 Villigen PSI, Switzerland}
\affiliation{Department of Quantum Matter Physics, University of Geneva, 
CH-1211 Geneva, Switzerland} 

\date{\today}

\begin{abstract}
The challenge of one-dimensional systems is to understand their physics 
beyond the level of known elementary excitations. By high-resolution neutron 
spectroscopy in a quantum spin ladder material, we probe the leading 
multiparticle excitation by characterizing the two-magnon bound state at 
zero field. By applying high magnetic fields, we create and select the 
singlet (longitudinal) and triplet (transverse) excitations of the fully 
spin-polarized ladder, which have not been observed previously and are close 
analogs of the modes anticipated in a polarized Haldane chain. Theoretical 
modelling of the dynamical response demonstrates our complete quantitative 
understanding of these states.
\end{abstract}


\maketitle

Throughout physics, one-dimensional (1D) systems show a range of intriguing 
and unconventional phenomena. The dominance of quantum fluctuations in 1D 
materials makes them the ultimate form of quantum matter. The physics of the 
ground states and elementary, low-energy excitations of 1D systems are by now 
rather well understood in theory \cite{Gia04} and have been realized in 
experiment in a number of quite different fields, including conducting wires 
\cite{rcw}, atomic chains \cite{rac}, quantum magnets \cite{Mik04}, and 
ultracold atoms \cite{rua}. Looking forwards, the next frontier is to 
understand and control the physics of these systems on all energy scales, 
including their multiparticle excitations and topological states.

Quantum magnets provide an excellent arena not only for quantitative 
measurements of the strongly correlated quantum wave function but also for 
its systematic control by applied external parameters \cite{rgrt,War13,
Mer14}. Among the systems whose elementary magnetic excitations are 
already well characterized, one key model is the $S = 1/2$ ``ladder,'' 
consisting of two coupled spin chains \cite{rdr}, with detailed 
experimental studies performed on genuine ladder materials including 
La$_4$Sr$_{10}$Cu$_{24}$O$_{41}$ \cite{Not07}, BPCB \cite{Pat90,Wat01,Rue08,
Kla08,Lor08,Thi09a,Thi09b} and DIMPY \cite{Sha07,Hon10,Sch12,Sch14}. The 
ladder has many parallels to another cornerstone model, the Haldane ($S = 1$) 
chain \cite{Hal83,Aff87}, and significant progress has been made in calculating 
their dynamical response \cite{Bou11}. However, genuine Haldane materials with 
accessible energy scales have proven difficult to find \cite{Zhe03,Ber15,Zap06}, and thus the response in strong magnetic fields remains an open problem \cite{Kol02,Ess04}.

In this Letter, we report on measurements of single- and multi-magnon 
excitations in the spin ladder bis-piperidinium copper tetrachloride (BPCC). 
By exploiting the elegant parity selectivity of the ladder geometry, at zero 
field we demonstrate the presence of a strong two-magnon bound-state triplet 
over half of the Brillouin zone and quantify its spectral weight. At high 
fields, we demonstrate the selection criteria for the singlet excitation, 
or amplitude mode, of the fully field-polarized (FP) phase, as well as for 
its triplet (transverse) mode. Both are unknown in a conventional ferromagnet 
and have direct analogs in the FP Haldane chain. By detailed analytical and 
numerical modelling, we describe our intensity measurements with quantitative 
accuracy. 

The spin ladder has two basic magnetic interactions, the rung ($J_r$) and  
leg ($J_l$) couplings, and one ratio, $\gamma = J_l/J_r$. BPCB ($\gamma \simeq 
0.26$) and DIMPY ($\gamma \simeq 1.7$) exemplify contrasting regimes of 
ladder behavior. The chloride analog of BPCB, (C$_5$D$_{12}$N)$_2$CuCl$_4$ 
(BPCC, Fig.~S2) crystallizes in the monoclinic space group P$2_1$/c.
Two halide bridges between pairs of Cu$^{2+}$ ions form the rung dimer ($J_r$). 
A further halide bridge between ions repeating periodically in ${\hat a}$ 
provides $J_l$. BPCC is an exceptional realization of a very clean $S = 1/2$ 
spin ladder with, as we will show, three attributes ideal for our studies. 
First, the ladders are well separated by piperidinium groups, making them 
effectively isolated. Second, the ratio $\gamma \simeq 0.4$ \cite{Ryl14} is 
perfectly suited for observing bound states in the two-magnon sector (whose 
weight scales with $\gamma^2$), without strong interference from the scattering 
states. Third, the low energy scales in BPCC \cite{Taj04,Ryl14} allow one to 
work well within the FP phase at laboratory magnetic fields.

High-quality deuterated single-crystal samples were synthesized and five 
crystals of total mass 2.4 g were coaligned on the MORPHEUS instrument at 
SINQ for the collection of inelastic neutron scattering data on the 
time-of-flight spectrometer LET at ISIS \cite{Bewley2011p1}. The sample 
was mounted inside a 9~T vertical cryomagnet, on a dilution insert with a 
stable base temperature of 60~mK. Frame-rate multiplication was used with 
a primary incoming neutron energy of 2.5~meV. Data at zero field were 
collected at 104 sample rotation angles and processed using the MANTID 
program \cite{mantid}. The resulting datasets for the dynamical structure 
factor, $S(\mathbf{Q},\omega)$, were analyzed with the HORACE software 
package \cite{horace}. 

\begin{figure}[p]
\centering
\includegraphics[width=\linewidth]{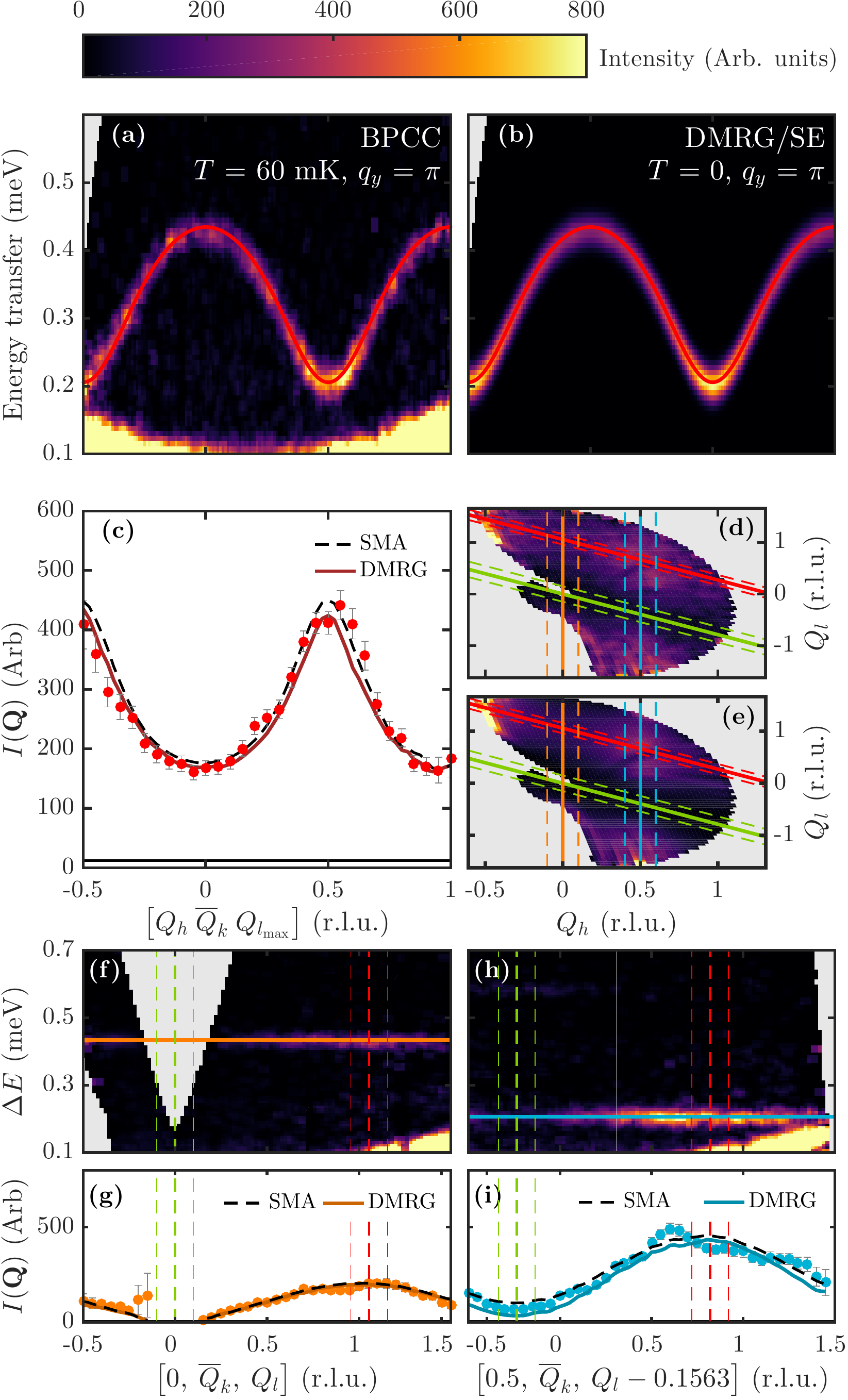}
\caption{{\bf One-magnon excitations in BPCC.} (a) $S(\mathbf{Q},\omega)$ 
measured in the $q_y = \pi$ sector. (b) $S(\mathbf{Q},\omega)$ obtained 
by DMRG and series-expansion (SE) calculations, whose results are 
indistinguishable. The solid red line is the dispersion obtained from 
strong-coupling [Eq.~(S5)] and bond-operator treatments [Eq.~(S6)]. (c) 
$I(\mathbf{Q})$ obtained by integration over $Q_k$ (denoted ${\overline Q}_k$) 
and over the energy range $0.19 \le \omega \le 0.43$ meV, shown as red points 
and computed in the SMA (dashed black) and by DMRG (solid red lines); the 
solid black line denotes the background and $Q_{l_{\rm max}}$ is defined in 
Eq.~(S4). (d) Measured structure factor and (e) that calculated by DMRG; 
${\bf Q}$ vectors for the ladder direction in the $q_y = \pi$ and $q_y = 0$ 
sectors are shown respectively as red and green solid lines, with those 
transverse to the ladder at $Q_h = 0$ and 0.5 shown respectively in blue 
and orange; dashed lines mark the transverse integration ranges. (f) 
$S({\bf Q},\omega)$ as a function of $Q_l$, taken at $Q_h = 0$ and cutting 
the $q_y = \pi$ [intensity maxima, red dashed lines, panel (a)] and $q_y = 0$ 
[minima, green, Fig.~\protect{\ref{fig:st_tp}}(a)] sectors. The solid orange 
line is the one-magnon band maximum [Eq.~(S5)]. (g) $I({\bf Q})$ ($0.40 \le 
\omega \le 0.43$ meV, orange points), compared with SMA (dashed black) and 
DMRG calculations (solid orange lines). (h) $S({\bf Q}, \omega)$ measured for 
scattering vectors [0.5 ${\overline Q}_k$ $Q_l - 0.1563$] at the band minimum. 
(i) $I({\bf Q})$ ($0.19 \le \omega \le 0.23$ meV, blue points), with 
corresponding theoretical predictions. 
\label{fig:st_zero}}
\end{figure}

In the two-leg ladder, the rung-singlet wave function is antisymmetric 
whereas the triplets are symmetric (represented schematically in Fig.~S1). 
The geometry therefore allows a full parity selection between odd ($q_y = 
\pi$; e.g.~singlet-triplet) and even ($q_y = 0$; e.g.~triplet-triplet) 
excitation processes \cite{rtu,Bou11}, where $q_y$ is the wave vector 
across the ladder unit. There is no mixing between modes of opposite parity 
and $S(\mathbf{Q},\omega)$ separates completely into odd- and even-parity 
sectors, whose intensity maxima occur in very different parts of the Brillouin 
zone [Figs.~\ref{fig:st_zero}(d) and \ref{fig:st_zero}(e)]. Although the 
crystallography of BPCC dictates that there are two different ladder 
orientations, this parity selection remains possible \cite{rtu,Bou11}, 
as shown in Sec.~S1 of the supplemental material (SM) \cite{SuppM}. 

We begin our presentation at zero applied field. We refer to the elementary 
one-triplet excitation of the ladder as a ``magnon.'' The one-magnon 
dispersion relation is clearly evident in Fig.~\ref{fig:st_zero}(a), where 
the four-dimensional (4D) intensity dataset is analyzed to select the 
$q_y = \pi$ sector. Our data treatment is discussed in Sec.~S2 of the SM. 
We observe an excitation with gap 0.2 meV, bandwidth 0.23 meV, and 
periodicity 2$\pi$, suggesting that $J_r$ is significantly stronger than 
$J_l$. 

Several theoretical approaches give good descriptions of the two-leg 
ladder in the strong-rung regime. From an optimized fit of the one-magnon 
data to a strong-coupling expansion \cite{Rei94} [Eq.~(S5) of Sec.~S3], 
we deduce the exchange parameters $J_r$ = 0.295(8) meV and $J_l$ = 0.116(1) 
meV; the resulting dispersion is shown as the red lines in 
Figs.~\ref{fig:st_zero}(a) and \ref{fig:st_zero}(b). The corresponding 
values from the bond-operator technique \cite{rnr} [Eq.~(S6)] are $J_r$
 = 0.306(8) meV and $J_l$ = 0.113(1) meV, and the dispersion is identical 
within the error bars. Thus the leg exchange in BPCC, with $\gamma 
\simeq 0.39$, is more significant than in BPCB, which is the key to our 
quantitative observations of bound states. We perform both DMRG and 
systematic high-order series-expansion calculations for $S(\mathbf{Q},\omega)$ 
(Sec.~S3), which capture the contributions of all excitation channels; with 
optimized parameters $J_r$ = 0.295 (0.294) meV and $J_l$ = 0.115 (0.117) meV 
for DMRG (series expansions), both yield excellent agreement with 
the one-magnon measurements, as shown in Fig.~\ref{fig:st_zero}(b).

For a quantitative discussion of the scattering intensity, we integrate 
$S(\mathbf{Q},\omega)$ over specific energy ranges and consider the 
resulting structure factor, $I({\bf Q})$. The quantity $I(\mathbf{Q})$ 
obtained by integrating over the full one-magnon energy range ($0.19 \! 
\le \! \omega \! \le \! 0.43$ meV) is shown in Fig.~\ref{fig:st_zero}(c). 
Our DMRG (solid line) and series-expansion calculations reproduce not only 
the dispersion but also the spectral weight to very high accuracy. To analyze 
the one-magnon intensity, we exploit the dominance of this mode in the 
spectrum to apply the single-mode approximation (SMA, Sec.~S3). With prior 
knowledge of $J_r$ and $J_l$, the average spin correlations, $\langle {\bf S}_1 
\! \cdot {\bf S}_2 \rangle$, can be extracted separately for the rung and leg 
bonds \cite{rnr}. Here, however, the unknown overall scale factor restricts 
us to deducing their ratio \cite{rtu}, whose expected theoretical value 
(Sec.~S3) is 0.26(1). From our data, we obtain a correlation ratio of 
0.22(1) and the fit to $I({\bf Q})$ shown by the dashed line in 
Fig.~\ref{fig:st_zero}(c). Thus our one-magnon intensity measurements both 
quantify the correlations (entanglement) in the quantum wave function and set 
the scale factor for intensity calculations in all other scattering sectors. 

The intensity distribution for wave vectors transverse to the ladder 
direction, shown in Figs.~\ref{fig:st_zero}(f) to \ref{fig:st_zero}(i), probes 
the diagonal ($J_d$) and interladder ($J'$) interactions. We find that the 
one-magnon mode is completely non-dispersive within the instrumental 
resolution, placing (Sec.~S2) an upper bound of 0.006(1) meV on any possible 
couplings of these types. While this is a maximum of 2\% of $J_r$, in fact a 
far more stringent bound on $J'$ is provided by the absence of magnetic order 
down to 35 mK, making the ladders in BPCC at least as isolated as those in 
BPCB, where $J'/J_r \approx 0.2$\% \cite{Thi09b}. 

\begin{figure}[t]
\centering 	
\includegraphics[width=\linewidth]{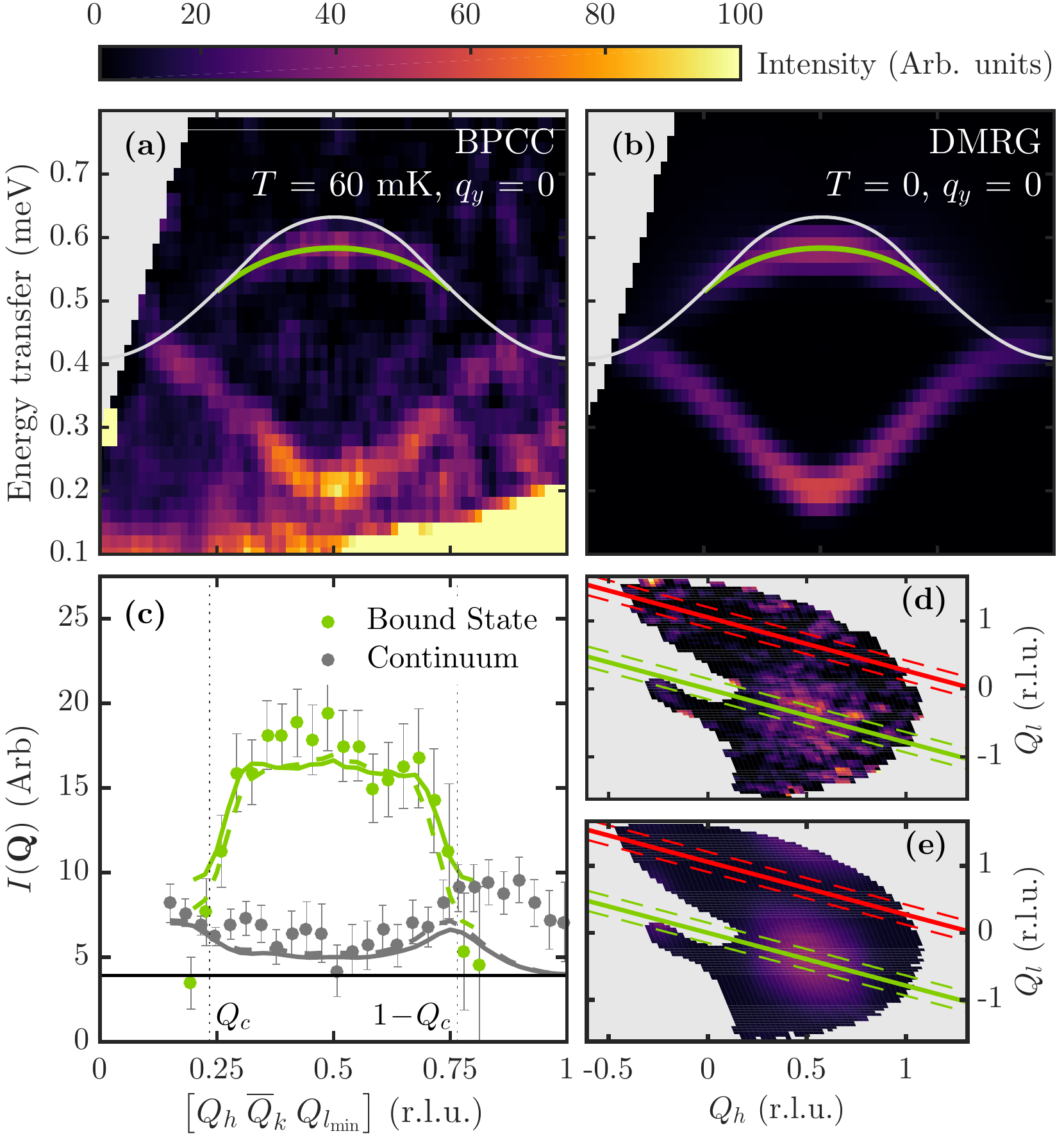}
\caption{{\bf Two-magnon excitations in BPCC.} (a) $S({\bf Q},\omega)$ 
measured for $q_y = 0$. The green line indicates the dispersion [Eq.~(S8)] and 
range [Eq.~(\ref{eqn:qlb})] over which the triplet component of the two-magnon 
bound state is expected in BPCC. The white line marks the boundary of the 
two-magnon continuum [Eq.~(S9)]. (b) Corresponding DMRG calculations. (c) 
Bound-state (green) and continuum (grey) contributions to $I({\bf Q})$, 
obtained by integrating over energy windows below and above the two-magnon 
continuum edge; measured intensities (points) are compared with DMRG (solid 
line) and series-expansion (dashed line) calculations. The black line 
denotes the background and $Q_{l_{\rm min}}$ is defined in Eq.~(S4). Measured 
(d) and calculated (e) structure factors for the $q_y = 0$ sector in the 
$(Q_h,{\overline Q}_k,Q_l)$ plane; red and green lines as in Fig.~1.
\label{fig:st_tp}}
\end{figure}

Next we consider multi-magnon excitation processes, whose intensity 
contributions we separate unambiguously by exploiting the parity of 
the ladder structure (Sec.~S1). The symmetric ($q_y = 0$) channel in 
Fig.~\ref{fig:st_tp} is the maximum of all excitation processes changing 
the triplet count by an even number and is dominated by two-magnon excitations 
into both continuum states and bound states. The zero-field spectrum of DIMPY 
has been found to exhibit both features \cite{Sch12}. However, in BPCC the 
continuum contributions are extremely weak, allowing the dispersion, spectral 
weight, and termination of the two-magnon bound state to be measured in a 
degree of detail extremely difficult to achieve either in DIMPY or in the 
dimerized-chain material Cu(NO$_3$)$_2$$\cdot$2.5D$_2$O \cite{Ten03}.

$S({\bf Q},\omega)$ measured with $q_y = 0$ is shown in 
Fig.~\ref{fig:st_tp}(a). We observe a clear, discrete mode lying at an energy 
of 0.58 meV at $Q_h = 0.5$, below the continuum edge, and dispersing downwards 
before disappearing into the continuum, whose contribution is indiscernible in 
Fig.~\ref{fig:st_tp}(a). From the selection rules for neutron scattering, this 
mode is the triplet branch ($S_{\rm tot} = 1$) of the two-magnon bound state; 
the singlet ($S_{\rm tot} = 0$) branch is detectable by light scattering 
\cite{Win01}, and both singlet and quintet ($S_{\rm tot} = 2$) branches by 
neutron spectroscopy at higher temperatures \cite{rhmn}. We note that 
one-magnon spectral weight is detectable in this sector because the geometry 
of BPCC excludes perfect destructive interference (Sec.~S1). However, in 
contrast to the discussion of magnon ``termination'' \cite{Sto06}, here 
the one- and two-magnon excitations are prevented by their opposite 
parities from mixing where they overlap in energy [as can be seen in 
Fig.~\ref{fig:st_zero}(a)].

The bound state overlaps with the two-magnon continuum below a critical 
wave vector, $Q_c$, and decays very readily in this range \cite{Zhe01}. 
It is therefore clearly visible only when $Q_c < Q_h < 1 - Q_c$, with 
\begin{equation} 
\label{eqn:qlb}
Q_c = {\textstyle {\frac{1}{3}}} - {\textstyle \frac{5}{4\sqrt{3}\pi}} \gamma
 - {\textstyle \frac{109}{96\sqrt{3}\pi}} \gamma^2 + \mathcal{O}(\gamma^3)
\end{equation}
in a strong-coupling expansion (Sec.~S3). For BPCC, Eq.~(\ref{eqn:qlb}) with 
$\gamma = 0.39$ leads to $Q_c = 0.212$. However, this value of $\gamma$ lies 
well in the regime where high orders are required for an accurate description. 
A range of $0.235 \le Q_h \le 0.765$ is computed in Ref.~\cite{Zhe01} by 
working to 12th order in $\gamma$, and in Fig.~\ref{fig:st_tp}(b) we require 
the 10th-order dispersion of the two-magnon bound state \cite{Kne04} and the 
continuum edge. From our DMRG and series-expansion calculations, we 
find that both the dispersion [Figs.~\ref{fig:st_tp}(b) and S3] and the 
scattered intensity [Figs.~\ref{fig:st_tp}(c), \ref{fig:st_tp}(d), and 
\ref{fig:st_tp}(e)] of both the bound-state mode and the two-triplet 
continuum agree exceptionally well with the data. Thus our results also 
quantify the termination of the bound state [Fig.~\ref{fig:st_tp}(c), 
Sec.~S3].

\begin{figure}[t]
\centering 	
\includegraphics[width=\linewidth]{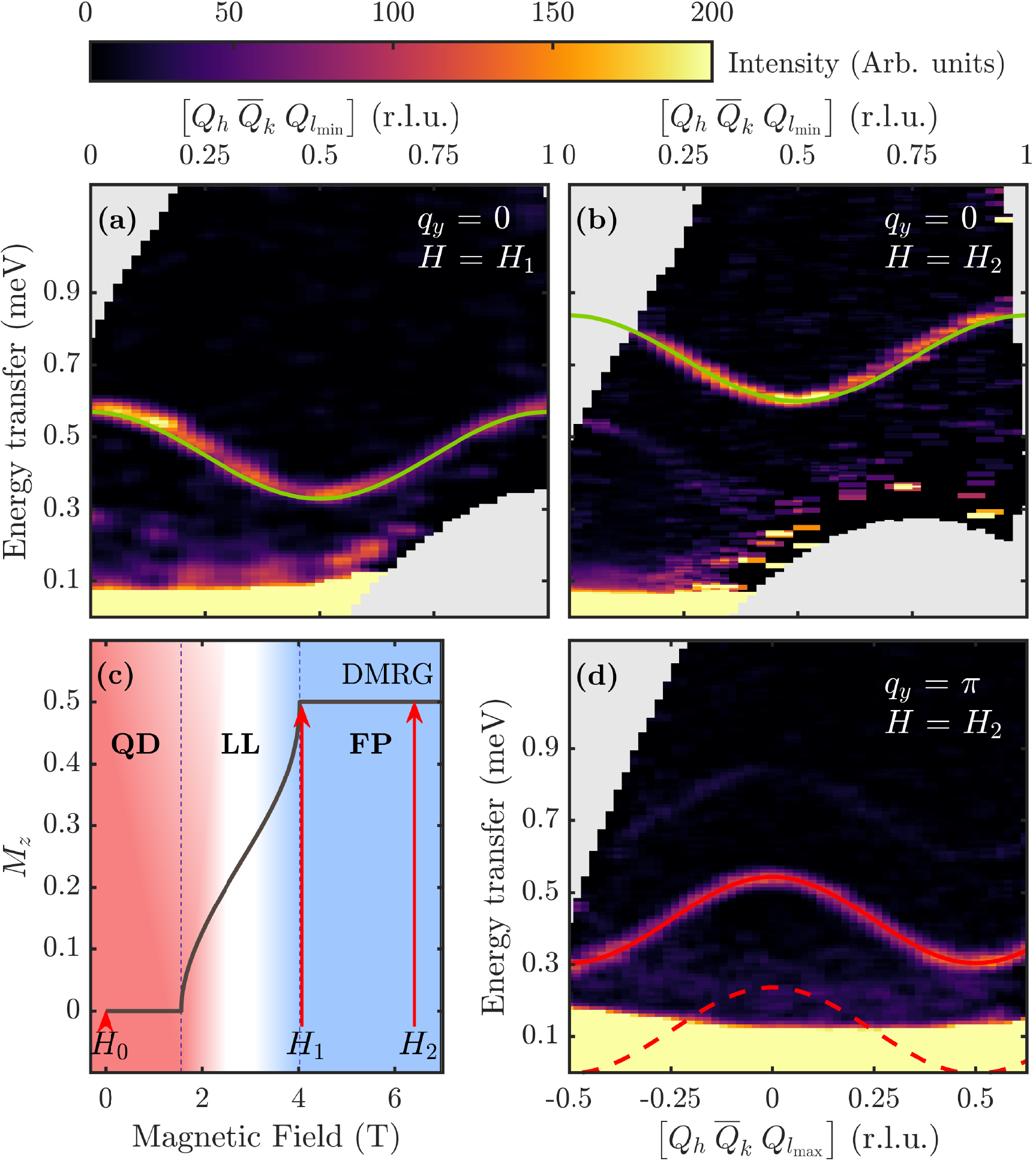}
\caption{{\bf Excitations in FP BPCC.} (a) $S({\bf Q},\omega)$ measured at 
$H_1 = 4.067$ T $\simeq H_s$ and (b) $H_2 = 6.141$ T, both in the $q_y = 0$ 
sector. (c) Magnetization of BPCC (from DMRG) superimposed on the 
field-induced phase diagram; QD denotes the quantum disordered phase. 
(d) $S({\bf Q},\omega)$ measured at $H_2$ in the $q_y = \pi$ sector. Green 
and red solid lines are respectively fits to the triplet ($q_y = 0$) and 
singlet ($q_y = \pi$) modes using Eq.~(\ref{eqn:FM}). The dashed line in 
panel (d) indicates the position of the singlet at $H_s$ (not measured), 
where the spectrum is dominated at low energies by incoherent elastic 
scattering.
\label{fig:FP_Dis}}
\end{figure}

We turn now to the excitation spectrum at high magnetic fields. Quantum 
magnets with low exchange energies exhibit field-induced critical points 
marking transitions between several types of exotic quantum phase in 
materials including Cs$_2$CuCl$_4$ \cite{Col02}, CuSO$_4$$\cdot$5D$_2$O 
\cite{Mou13}, and BPCB \cite{Rue08,Kla08,Thi09a}. For two-leg ladders, 
the physics of the intermediate-field, ``Luttinger-liquid'' (LL) regime 
[Fig.~\ref{fig:FP_Dis}(c)], which possesses a continuum of gapless, 
fractional spin excitations, has been studied in depth in BPCB and we 
do not repeat this analysis for BPCC here. 

However, in all ladder materials studied to date, the dynamical properties 
have been measured only below (DIMPY, $H_s \simeq 29$ T, \cite{Sch12,Sch14}) 
or at the saturation field (BPCB, $H_s$ = 14 T \cite{Thi09a}). The exchange 
parameters of BPCC place full saturation well within our instrumental 
capabilities. Here we present previously unavailable measurements of the 
spectrum of the FP ladder.

Figure \ref{fig:FP_Dis} shows the dynamical structure factor of FP BPCC. 
Parity remains a good quantum number at high fields and in each sector we 
find one mode. At $H_s \simeq H_1$ = 4.067 T, in the $q_y = 0$ sector we 
observe [Fig.~\ref{fig:FP_Dis}(a)] a mode with a purely sinusoidal dispersion 
and a gap of 0.3 meV ($J_r$). On increasing the field to $H_2$ = 6.141 T 
[Fig.~\ref{fig:FP_Dis}(b)], this gap increases linearly with $H$; at $H_2$ 
with $q_y = \pi$, we observe a mode with identical dispersion but a gap 
smaller by precisely $J_r$ [Fig.~\ref{fig:FP_Dis}(d)]. 

The FP ground state is one of ``up-up'' triplets, $|t_+ \rangle = 
|\!\! \uparrow \uparrow \rangle$, on every ladder rung. Parity 
selection separates triplet-singlet ($q_y = \pi$) from triplet-triplet 
excitations ($q_y = 0$). The excitations of the FP phase are quite unlike 
the modes of a ferromagnet, with or without an applied field, not least in 
that the band minimum remains at $q_h = \pi$. Although all FP excitations 
are $\Delta S_z = - 1$, this can be achieved with $\Delta S_{\rm tot} = 0$ 
or $\Delta S_{\rm tot} = - 1$. The ladder provides a transparent situation 
where the latter is simply the singlet excitation of a rung [Fig.~S5(a)] 
and the former the $t_0$ triplet [Fig.~S5(b)]. The bond-operator method 
is exact at $H \ge H_s$, where quantum fluctuations are completely quenched, 
giving (Sec.~S3)
\begin{equation}
\begin{aligned}
\omega_s (q) & = g \mu_B (H - H_s) + J_l (1 + \cos q), 
\\ \omega_0 (q) & = g \mu_B (H - H_s) + J_r + J_l (1 + \cos q),
\label{eqn:FM}
\end{aligned}
\end{equation}
with $H_s = (J_r + 2 J_l)/ g \mu_B$. The upper triplet mode, $t_-$, is 
invisible to neutron scattering and has no dispersion. 

The singlet excitation is a longitudinal, or amplitude, mode of the 
field-enforced order, and is gapless at $H_s$ but has a gap of $H - H_s$ 
above this field; being antisymmetric, it is found with $q_y = \pi$, as 
in Fig.~\ref{fig:FP_Dis}(d). This mode is also expected in the Haldane 
chain, albeit in the adapted form of leg singlets formed from the auxiliary 
$S = 1/2$ spins of the AKLT \cite{Aff87} description [Fig.~S5(c)]. The 
$\Delta S = 0$ excitation is the gapped triplet, $t_0$, a transverse spin 
mode also exactly analogous to the $t_0$ excitation of the FP Haldane 
chain [Fig.~S5(d)]. By its symmetric nature, it is the mode observed in 
Figs.~\ref{fig:FP_Dis}(a) and \ref{fig:FP_Dis}(b), lying higher than the 
singlet by energy $J_r$ [Eq.~(\ref{eqn:FM})]. This $t_0$ mode has not been 
observed in any other ladder or Haldane system; it is also expected in FP 
alternating-chain materials, such as Cu(NO$_3$)$_2$$\cdot$2.5D$_2$O 
\cite{Sto14}, although its interpretation is more complicated in the 
absence of parity selection. By simultaneous fits to Eq.~(\ref{eqn:FM}) 
at $H_1$ and $H_2$, we find that exchange parameters $J_r$ = 0.294(1) meV 
and $J_l$ = 0.120(1) meV, combined with a $g$ value of 2.26(1), provide an 
extremely accurate account of the data.

We have measured the magnetic excitations of the two-leg ladder BPCC 
at all wave vectors and magnetic fields. At zero field, we exploit the 
parity-selectivity of the ladder to achieve a total separation of the one- 
and two-triplet excitation sectors, finding the clearest known example of 
the two-magnon bound state. In the fully spin-polarized regime, we select 
the gapped singlet (amplitude) and triplet (phase) modes of the ladder. Thus 
we provide a systematic understanding of the magnetic excitations in a broad 
family of gapped 1D quantum magnets, including ladder, alternating-chain, and 
Haldane systems. For BPCC, we demonstrate that the magnetic response can be 
modelled with complete quantitative accuracy, in both dispersion and 
intensity, by modern DMRG and series-expansion techniques. 

This work was supported by the Swiss National Science Foundation, the 
German Research Society (DFG), and the Royal Society. It is based on 
experiments performed at the Swiss spallation neutron source SINQ at 
the Paul Scherrer Institute and at the UK spallation neutron source ISIS 
at the Rutherford-Appleton Laboratory.

\end{document}

\title{Supplemental Material for ``Bound states and field-polarized Haldane 
modes in a quantum spin ladder''}

\author{S. Ward}
\affiliation{Laboratory for Neutron Scattering and Imaging, Paul Scherrer 
Institut, CH-5232 Villigen PSI, Switzerland}
\affiliation{London Centre for Nanotechnology and Department of Physics and 
Astronomy, University College London, London WC1E 6BT, England, United Kingdom}
\affiliation{Department of Quantum Matter Physics, University of Geneva, 
CH-1211 Geneva, Switzerland} 

\author{P. Bouillot}
\affiliation{Department of Medical Imaging and Information Sciences, 
Interventional Neuroradiology Unit, University Hospitals of Geneva, 
CH-1211 Geneva, Switzerland}
\affiliation{Laboratory for Hydraulic Machines, \'{E}cole Polytechnique 
F\'{e}d\'{e}rale de Lausanne, CH-1015 Lausanne, Switzerland} 

\author{C. Kollath}
\affiliation{Department of Quantum Matter Physics, University of Geneva, 
CH-1211 Geneva, Switzerland} 
\affiliation{HISKP, University of Bonn, Nussallee 14-16, 53115 Bonn, Germany}

\author{T. Giamarchi}
\affiliation{Department of Quantum Matter Physics, University of Geneva, 
CH-1211 Geneva, Switzerland} 

\author{K. P. Schmidt}
\affiliation{Theoretische Physik I, Staudtstrasse 7, FAU Erlangen-N\"urnberg, 
91058 Erlangen, Germany}

\author{B. Normand}
\affiliation{Laboratory for Neutron Scattering and Imaging, Paul Scherrer 
Institut, CH-5232 Villigen PSI, Switzerland}

\author{K. W. Kr\"amer}
\affiliation{Department of Chemistry and Biochemistry, University of Bern, 
CH-3012 Bern, Switzerland} 

\author{D. Biner}
\affiliation{Department of Chemistry and Biochemistry, University of Bern, 
CH-3012 Bern, Switzerland} 

\author{R. Bewley}
\affiliation{ISIS Facility, Rutherford Appleton Laboratory, Chilton, Didcot, 
Oxford OX11 0QX, United Kingdom} 

\author{T. Guidi}
\affiliation{ISIS Facility, Rutherford Appleton Laboratory, Chilton, Didcot, 
Oxford OX11 0QX, United Kingdom} 

\author{M. Boehm}
\affiliation{Institut Laue Langevin, 6 rue Jules Horowitz BP156, 38024 
Grenoble CEDEX 9, France} 

\author{D. F. McMorrow}
\affiliation{London Centre for Nanotechnology and Department of Physics and 
Astronomy, University College London, London WC1E 6BT, England, United Kingdom}

\author{Ch. R\"uegg}
\affiliation{Laboratory for Neutron Scattering and Imaging, Paul Scherrer 
Institut, CH-5232 Villigen PSI, Switzerland}
\affiliation{Department of Quantum Matter Physics, University of Geneva, 
CH-1211 Geneva, Switzerland} 


\maketitle

\subsection{S1. Neutron Scattering in a Two-Leg Spin Ladder}    

\subsubsection{Parity Selection}

\begin{figure}[t]
\includegraphics[width=7.5cm]{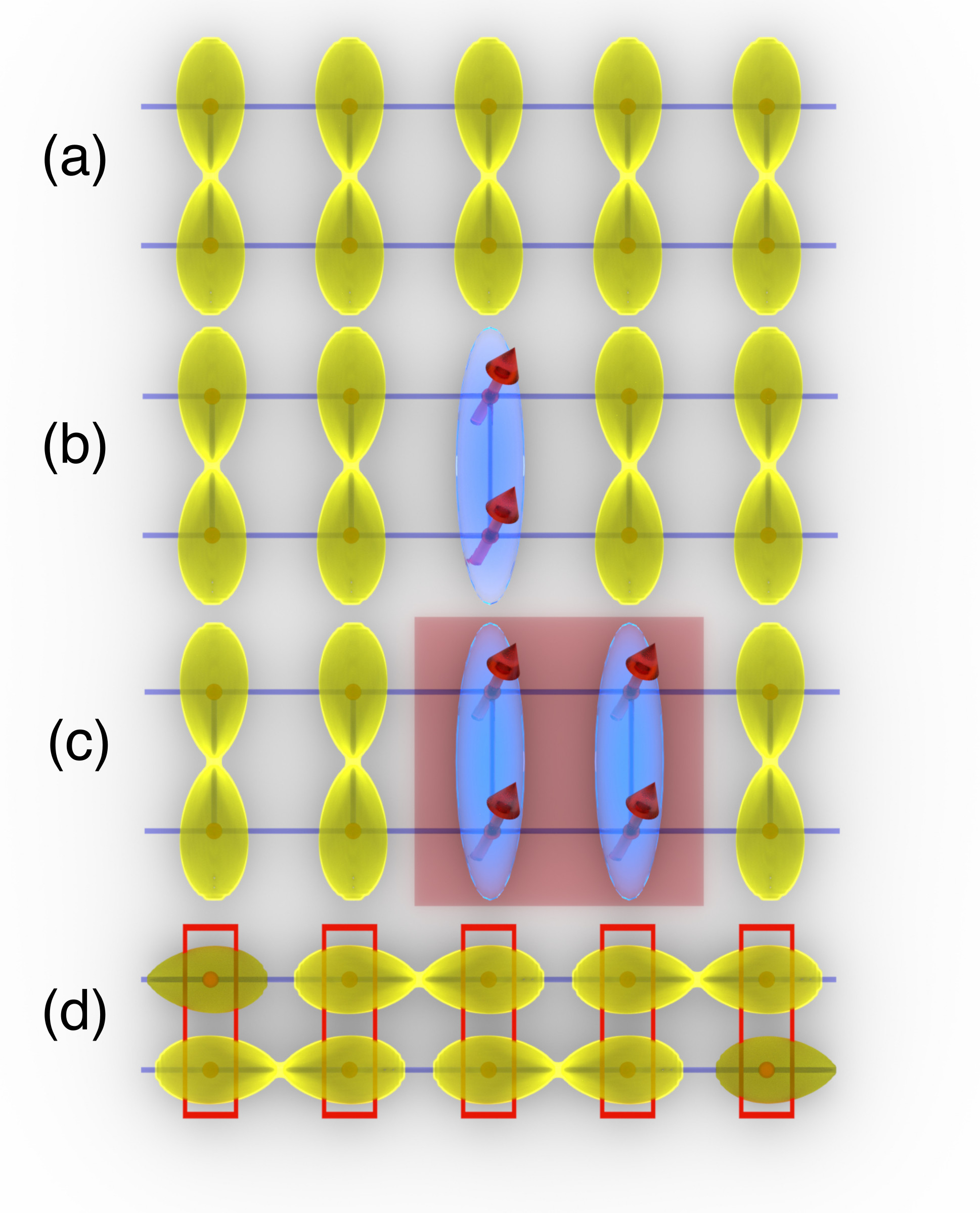}
\caption{\label{fsh0} Schematic representation of rung states 
and their symmetries in spin-ladder and Haldane-chain systems at zero field. 
(a) Ground state of antisymmetric rung singlets in the two-leg ladder. (b) 
One-magnon (symmetric, triplet) excited state. (c) Two-magnon excited state; 
the triplet pair may be in a net singlet, triplet, or quintet ($S_{\rm tot} = 0$, 
1, or 2) state but its parity is equal to that of the ground state and 
opposite to that of the one-magnon excitation. (d) Ground state of the 
Haldane chain in the AKLT representation \protect{\cite{Aff87}} of auxiliary 
$S = 1/2$ spin pairs at each site; the red boxes represent the operator 
projecting these spins to a real $S = 1$ state.} 
\end{figure}

A fundamental consequence of the two-leg geometry is that all excitations of 
the spin ladder have an exact even or odd parity. Excitations between singlet 
and triplet states have odd parity ($q_y = \pi$ sector) and may therefore be 
separated systematically from excitations of inter-triplet (or inter-singlet) 
character, which have even parity ($q_y = 0$ sector); this situation is 
represented schematically in Fig.~\ref{fsh0}. Most important for the present 
purposes is that one-magnon excitations are odd whereas two-magnon excitations 
are even; these opposite parities result in opposite phases for constructive or 
destructive interference, and hence in a complete separation of the maximal 
intensities of the two sectors in reciprocal space. A further consequence is 
that excitations of opposite parities do not mix, excluding the possibility 
of ``quasiparticle breakdown'' where they overlap.

\subsubsection{Crystal Structure of BPCC}

\begin{figure}[t]
\includegraphics[width=4.2cm]{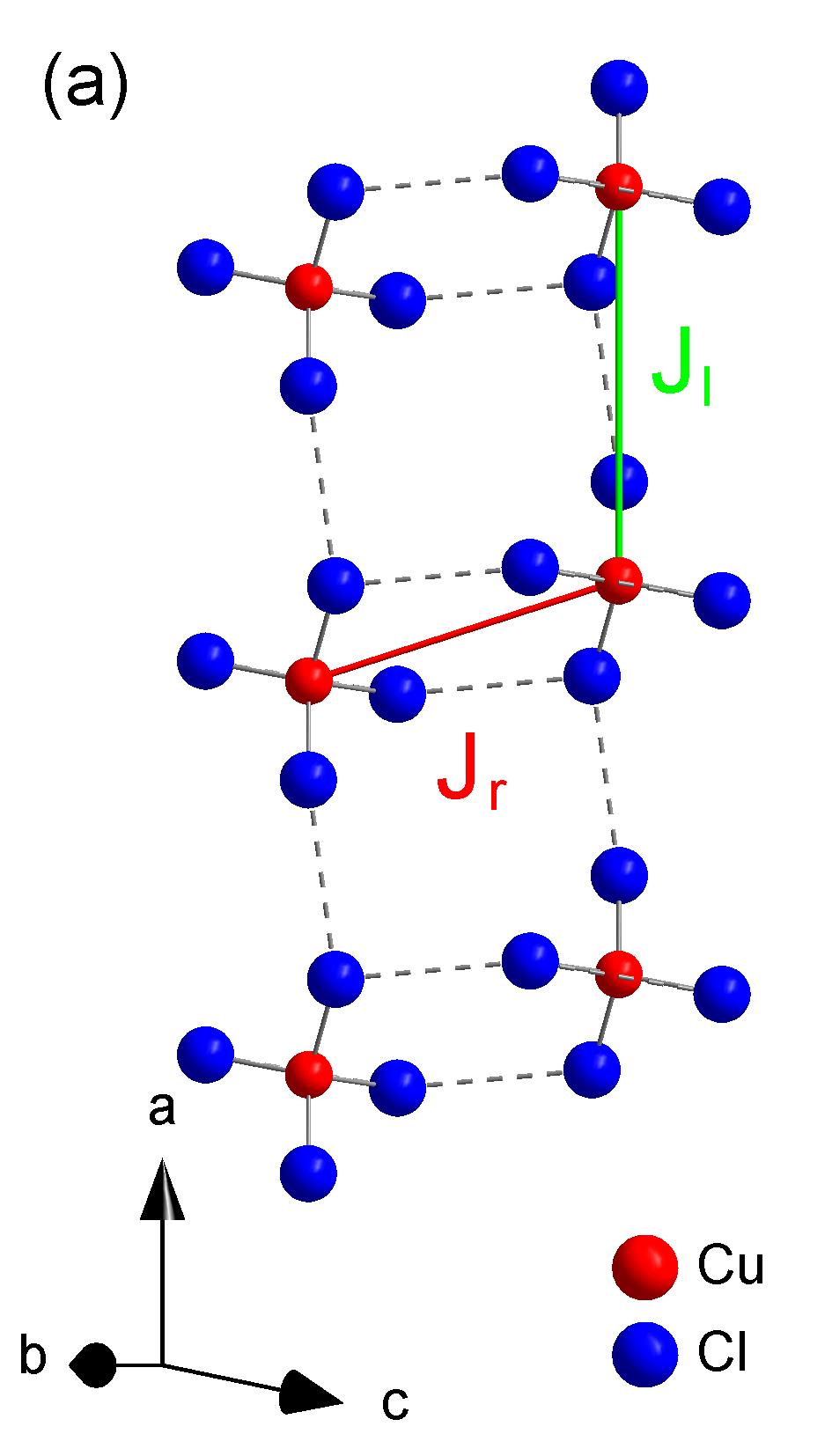}\hspace{0.1cm}\includegraphics[width=4.2cm]{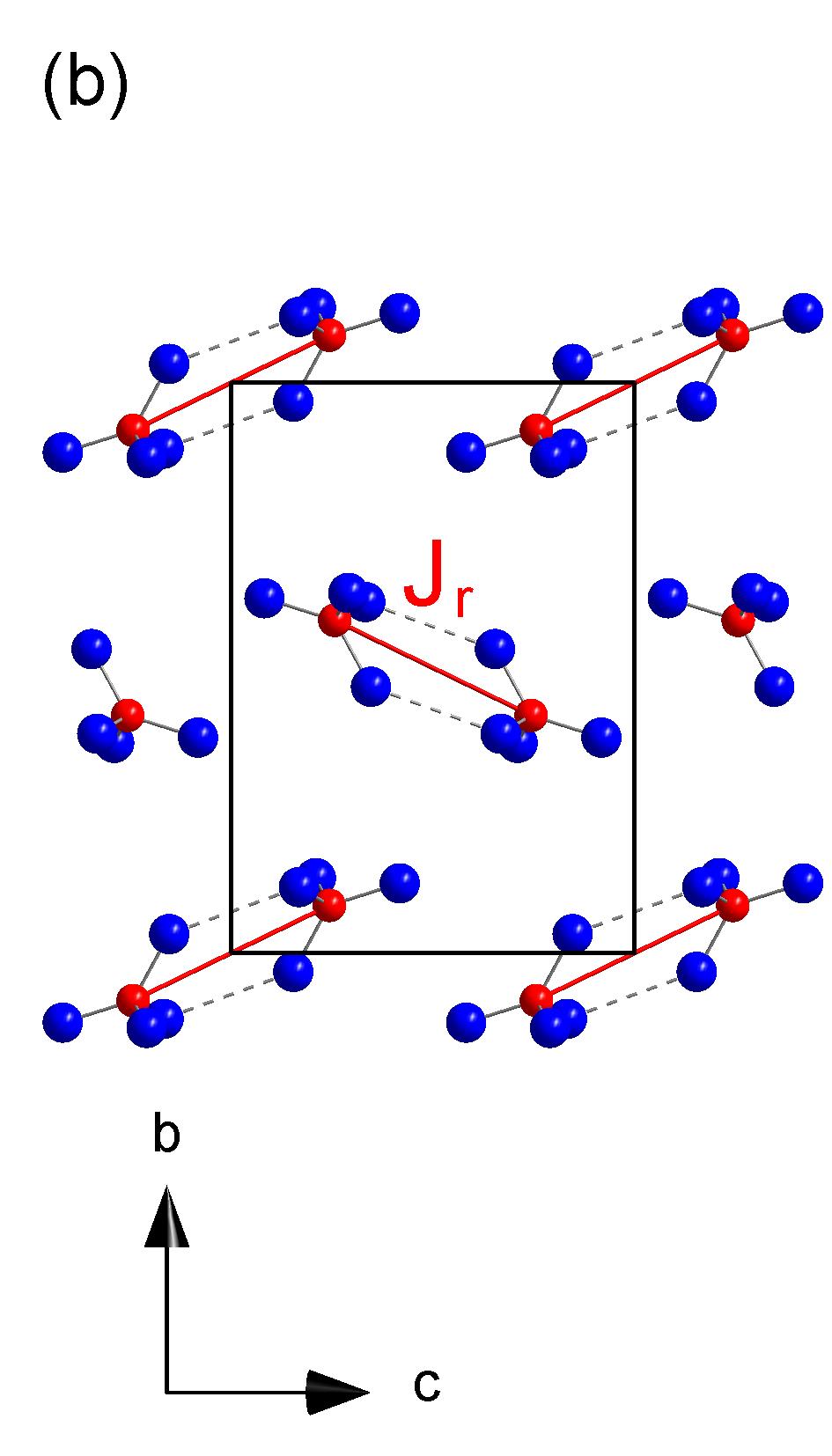}
\caption{\label{fstruc} Representation of the crystal structure of BPCC, 
highlighting (a) the positions of Cu and Cl atoms in a single ladder unit 
and (b) the relative positions of the different ladder units, which have two 
different orientations; the piperidinium cations between the ladders are 
omitted for clarity.}
\end{figure}

The structure of BPCC has the monoclinic space group P$2_1$/c, with lattice 
parameters determined by neutron diffraction at 1.6 K as $a = 8.08657(5)$ \AA, 
$b = 16.57703(9)$ \AA, $c = 12.14457(7)$ \AA, and $\beta = 102.2395(4)^\circ$ 
\cite{rwea}. The unit cell contains four Cu$^{2+}$ atoms ($S = 1/2$), which 
form two spin dimers constituting the ladder rungs, as shown in 
Fig.~\ref{fstruc}(a). These rung dimer units repeat periodically along the 
$\hat{a}$ direction, with an exchange interaction that forms the ladder legs 
[Fig.~\ref{fstruc}(a)]. Ladders neighboring in the $\hat{b}$ direction are 
related by a 2$_1$ screw axis and hence there are two types of spin ladder, 
identical (by symmetry) in their exchange interactions but different in 
orientation [Fig.~\ref{fstruc}(b)]. The inequivalent rung vectors are given by
\begin{equation}
\mathbf{r}_{1,2} = [0.3822 \; \pm 0.173 \: 0.4866];
\label{erv}
\end{equation}
in the same units, the leg vector is $[1 \; 0 \; 0]$. Despite this 
inequivalence, parity remains a good quantum number: excitations in the 
0 and $\pi$ sectors are associated with geometrical phase factors expressing 
their constructive or destructive interference, which determine the precise 
locations of maximal and minimal scattering intensity throughout the Brillouin 
zone for each sector. Because of the monoclinic structure of BPCC, these 
maxima and minima are determined not solely by the reciprocal-space component, 
$Q_h$, for the ladder direction, but are found respectively at $[Q_h \; 0\; 
Q_{l_{\rm max}}]$ and $[Q_h \; 0\; Q_{l_{\rm min}}]$, where $Q_{l_{\rm max}}$ and $Q_{l_{\rm 
min}}$ depend on $Q_h$, as we discuss below.

\subsubsection{Scattering Cross-Section}

The complete separation of excitations into even (0) and odd ($\pi$) parity 
sectors means that the neutron scattering cross section may also be decomposed 
into two separate types of contribution, $S^{\alpha\beta}_0$ and $S^{\alpha\beta}
_\pi$ \cite{rtu}. Although $\alpha$ and $\beta$ denote general spin indices, 
for pure Heisenberg interactions one has $S^{xx} = S^{yy} = S^{zz} = S^{+-} = 
S^{-+}$. When the two inequivalent wave vectors for the ladder rungs are taken 
into account, the total neutron scattering cross-section can be written as
\begin{eqnarray}
\!\! \frac{\mathrm{d}^2\sigma}{\mathrm{d}\Omega \mathrm{d}E'} & \propto & 4 
\! \left( \!\! 1 \! - \! \frac{Q_z^2}{Q^2} \! \right) \! [c ({\bf Q}) S_{0}^{zz} 
\! + \! s ({\bf Q}) S_{\pi}^{zz}] \label{eq} \\ & & + \!\! \left( \!\! 1 \! + \! 
\frac{Q_z^2}{Q^2} \! \right) \! [c ({\bf Q})\! (S_{0}^{+\!-} \!\! + \! S_{0}^{-\!+}) 
\! + \! s ({\bf Q}) \! (S_{\pi}^{+\!-} \!\! + \! S_{\pi}^{-\!+})], \nonumber
\end{eqnarray}
where $s(\mathbf{Q}) = \sum_{i=1,2} \sin^2 (\mathbf{Q \! \cdot \! r}_i/2)$ and  
$c(\mathbf{Q}) = \sum_{i=1,2} \cos^2 (\mathbf{Q \! \cdot \! r}_i/2)$ are phase 
factors and $Q_z$ is the out-of-plane wave-vector component. In practice, we 
focus directly on the dynamical structure factor, $S(\mathbf{Q},\omega)$, 
extracted from $\mathrm{d}^2\sigma/\mathrm{d}\Omega \mathrm{d}E'$.

Equation (\ref{eq}) may be used to quantify the statements made in the 
preceding subsections. The maximum in the structure factor contributed 
by one-magnon excitations is found by selecting the wave vectors ${\bf Q}$ 
maximizing $1 - \cos (\mathbf{Q \cdot r})$, where ${\bf r} = {\bf r}_1 - 
{\bf r}_2$ is the difference between the inequivalent rung vectors, i.e. 
${\bf Q}$ must satisfy 
\begin{equation}
2 \pi ({\bf Q \cdot r}) = (2n + 1) \pi,
\label{eqn:maxmin}
\end{equation}
where $n$ is an integer. Conversely, the minimum in the one-magnon structure 
factor, which coincides with the maximum in the two-magnon (even-parity) 
channel, is found when $n$ takes a half-integer value. One may then deduce 
that 
\begin{equation}
Q_l = [n + 1/2 - 0.3822 \, Q_h]/0.4866,
\label{eqn:maxmin2}
\end{equation}
with $n$ an integer for $Q_{l_{\rm max}}$ and $n$ half-integral for $Q_{l_{\rm min}}$.

\begin{figure}[t]
\includegraphics[width=\linewidth]{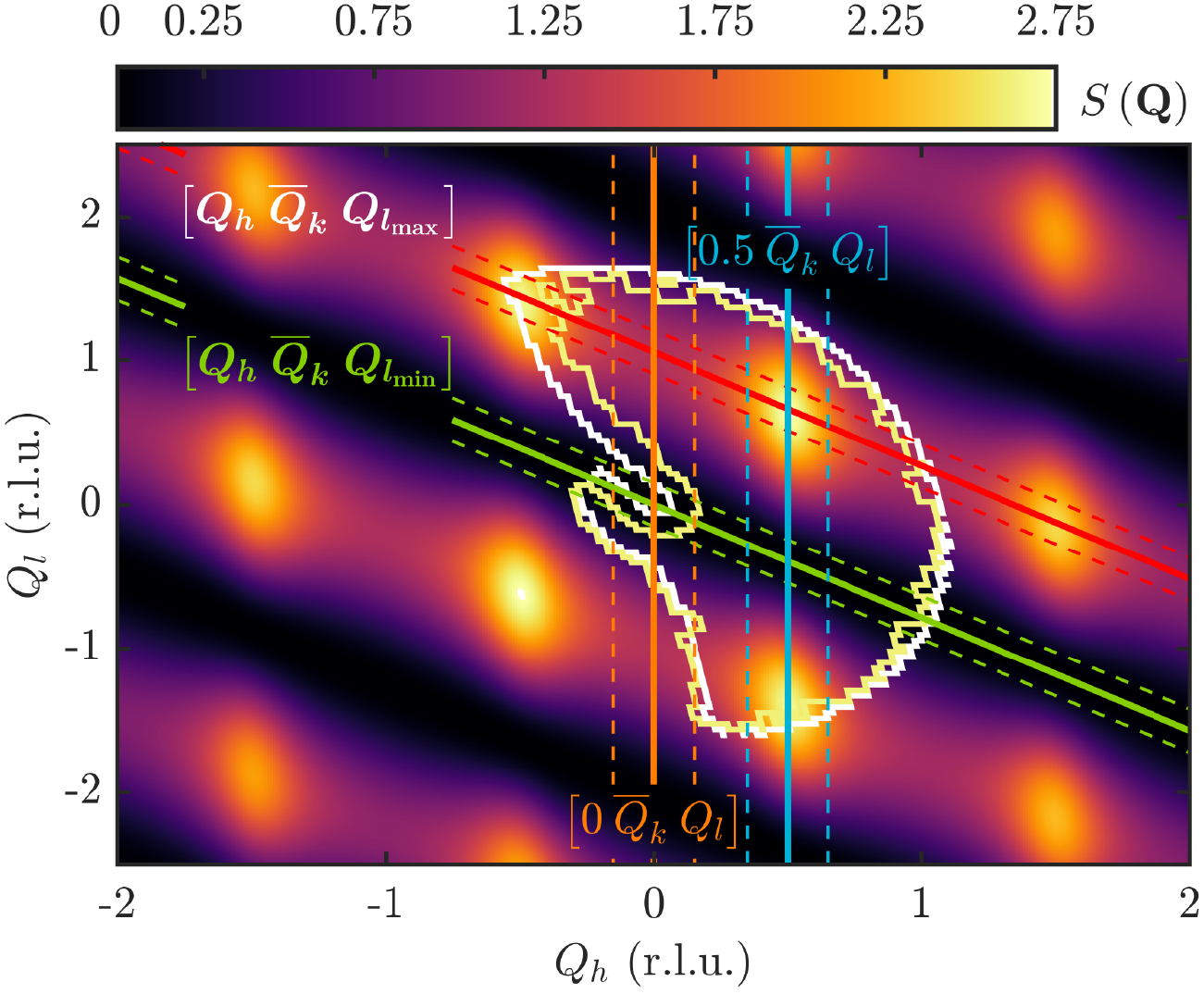}
\caption{\label{fig:QhQlRange} False-color representation of the simulated 
structure factor, $S(\mathbf{Q}, \omega)$, of the one-magnon excitation in 
BPCC, calculated in a strong-coupling analysis and using the single-mode 
approximation (Sec.~S3). The scattered intensity is shown in the 
horizontal scattering plane, $(Q_h,0,Q_l)$, for the energy value $\omega = 
\omega({\bf Q})$ of the one-magnon dispersion and for $Q_k = 0$. The white 
line marks the full extent of the available data in $Q_h$ and $Q_l$, which 
is determined by the detector coverage at the energy of the one-magnon mode. 
At the (higher) energy of the two-magnon continuum, the detector coverage has 
shrunk and is shown in yellow. The other lines are described in the text.}
\end{figure}

This information is represented in Fig.~\ref{fig:QhQlRange}, which shows 
the simulated structure factor of the one-magnon excitation, i.e.~the 
quantity $S(\mathbf{Q}, \omega({\bf Q}))$. The calculation of the dispersion 
relation, $\omega({\bf Q})$, and of the corresponding intensity is deferred 
to Sec.~S3. The red and green solid lines mark respectively the lines of 
maxima and minima of the one-magnon excitations of the spin ladder, whose
dynamical structure factor [$S(\mathbf{Q}, \omega)$] is shown in Fig.~1(a) 
of the main text. As noted above, the one-magnon minimum is the two-magnon 
maximum, for which $S(\mathbf{Q}, \omega)$ is shown in Fig.~2(a). Blue solid 
lines mark the maxima of the one-magnon excitations transverse to the ladder 
($Q_l$), which are obtained at the zone boundary in $Q_h$ and the spin-gap 
energy in $\omega$, and are studied in Fig.~1(h); orange lines mark the 
minima, which are found at the zone center and the band maximum [Fig.~1(f)]. 

\subsection{S2. Neutron Scattering Intensity Analysis}

Experiments were performed on the time-of-flight spectrometer LET, with the 
BPCC sample mounted in the (horizontal) \textit{ac} scattering plane. Data at 
zero applied magnetic field were collected for 104 rotation angles, corrected 
for detector efficiency and scattered-to-incident wave-vector ratio, $k_f/k_i$, 
using the MANTID program \cite{mantid}, and combined into a ``four-dimensional''
(4D, meaning three spatial dimensions and one energy) dataset using the 
Matlab-based HORACE analysis code \cite{horace}. The extent of the resulting 
dataset in reciprocal-space dimensions $Q_h$ and $Q_l$ is represented by the 
white line in Fig.~\ref{fig:QhQlRange}. For all of the results presented in 
the main text, the data were integrated over the $Q_k$ direction, representing 
out-of-plane scattering, where all modes are entirely non-dispersive; we refer 
to the results of this integration as a 3D dataset and denote this treatment 
of the $Q_k$ direction by ${\overline Q}_k$. Further integration over 
directions along or across the ladder was performed to obtain the final 
datasets; the dashed lines in Fig.~\ref{fig:QhQlRange} mark these 
transverse integration ranges. In all cases, the integration range was $\pm 
0.125$ r.l.u., a value achieving an acceptable signal-to-noise ratio while 
preserving mode coherence and sector separation.

For all quantitative analyses of the scattered intensity, we integrate 
$S({\bf Q}, \omega)$ over a chosen window of energy to obtain the structure 
factor, $I({\bf Q})$. In addition to improving very significantly the 
signal-to-noise ratio, this procedure also allows us to select the energy 
range in such a way as to remove as much as possible of the contamination 
due to elastic and multiple scattering events. This last includes scattering 
from the cryomagnet, which is responsible for the parabolic intensity visible 
at low energies in Figs.~1(a) and 2(a). The remaining background, due to 
detector noise and incoherent phonon scattering, is marked by the solid black 
lines in Figs.~1(c) and 2(c) of the main text. Fits to the models and numerical 
calculations were optimized by a least-squares approach. 

\subsubsection{Diagonal and Interladder Interactions} 

Exchange interactions between sites in neighboring spin ladders are 
reflected in the dispersion of the one-magnon mode for ${\bf Q}$ vectors 
transverse to the ladder axis. Any asymmetric exchange interactions within 
the ladder plaquettes, to which we refer here as ``diagonal,'' would also 
be manifest as a periodic modulation of the intensity distribution for the 
same ${\bf Q}$ vectors. Figure 1(f) shows $S({\bf Q},\omega)$ along [0 0 
$Q_l$], where the mode is clearly non-dispersive to within the resolution 
of our measurements. The red line is a theoretical result (Sec.~S3) with an 
interladder coupling of zero, and in fact the fitting process places an upper 
bound on interladder exchange of 0.006(1) meV. The corresponding structure 
factor, Fig.~1(g), is very well reproduced by calculations performed in the 
SMA and with DMRG. For transverse wave vectors chosen at the one-magnon gap, 
the dispersion is again completely flat [Fig.~1(h)] and the structure factor 
is described perfectly by a system of pure and isolated ladders [Fig.~1(i)]. 
Again there is no discernible periodicity, in dispersion or intensity, within 
the instrumental resolution, placing the same limit, 0.006(1) meV, on any 
possible diagonal or interladder interactions.

\subsection{S3. Theoretical Analysis of Strong-Rung Ladders}

\subsubsection{One-Magnon Dispersion} 

Several theoretical approaches are known to give good descriptions of the 
two-leg ladder in the strong-rung regime, including a direct perturbative 
expansion \cite{Rei94} and the bond-operator formalism \cite{rnr}. 
Numerically, both exact diagonalization (ED) and DMRG work well because of 
the short correlation length of the well-gapped system \cite{Rue08}. In 
terms of the parameter $\gamma = J_l/J_r$, to third order in a perturbative 
expansion one obtains the dispersion relation \cite{Rei94}
\begin{eqnarray}
\frac{\omega(q)}{J_r} & = & 1  + \frac{3}{4} \gamma^2 + \frac{3}{8} \gamma^3 
 + \left( \! \gamma - \frac{1}{4} \gamma^3 \! \right) \! \cos q \nonumber 
\\ & & - \frac{1}{4} \! \left( \gamma^2 + \gamma^3 \right) \cos 2 q 
+ \frac{1}{8} \gamma^3 \cos 3 q
\label{eqn:Res_Dis}
\end{eqnarray}
for the elementary triplet (one-magnon) excitation, where we have defined 
$q = 2 \pi Q_h$. In the bond-operator description, the dispersion is given by 
\begin{equation}
\omega(q) = \sqrt{({\textstyle \frac{1}{4}} J_r - \mu)({\textstyle \frac{1}{4}}
 J_r - \mu + 2 J_l {\overline s}^2 \cos q)},
\label{ebotd}
\end{equation}
where the $\mu$ is the chemical potential for the triplet excitations, which
are hard-core bosons, and ${\overline s}$ expresses the extent to which the 
ground state is one of pure rung singlets \cite{rnr}. 

\subsubsection{Single-Mode Approximation} 

While Eq.~(\ref{eq}) is an expression for the total neutron scattering 
cross-section, in many gapped quantum magnets the overwhelming majority 
of the scattered intensity can be found in the one-magnon branch of the 
excitation spectrum. In this situation, a meaningful analysis of the spectral 
weight is obtained within the single-mode approximation (SMA) \cite{Xu00}. 
Quite generally, the first-moment sum rule relates the integrated spectral 
weight of a spin ladder to the equal-time correlation functions in the form 
\begin{eqnarray}
\!\!\!\! \int \!\! d\omega & & \, \hbar \omega S(\mathbf{Q},\omega) = - 
{\textstyle \frac{2}{3}} (J_r \langle \mathbf{S}_i^m \! \cdot \mathbf{S}_i^n 
\rangle [1 - \cos (\mathbf{Q \! \cdot d})] \\ & & + J_l (\langle \mathbf{S}_i^m 
\! \cdot \mathbf{S}_{i+1}^m \rangle + \langle \mathbf{S}_i^n \! \cdot 
\mathbf{S}_{i+1}^n \rangle) [1 - \cos (\mathbf{Q \! \cdot r}_l)]) \nonumber, 
\end{eqnarray} 
where $m$ and $n$ denote the two chains of the ladder, $\mathbf{d} = 
\mathbf{r}_m - \mathbf{r}_n$ is the dimer (rung) bond vector and 
${\mathbf{r}}_l = \mathbf{r}_{m,i+1} - \mathbf{r}_{m,i}$ is the leg 
bond vector. In the SMA, the integral over $\omega$ includes a 
$\delta$-function, $\delta (\omega - \omega({\bf Q}))$, which transforms 
the left-hand side into a simple product of $\omega({\bf Q})$ and the 
structure factor, $I({\bf Q})$. The right-hand side is a set of ``bond 
energies,'' weighted by cosine factors determined from the geometry of 
the system. These bond energies are the product of the exchange interactions 
for each bond with the average bond spin correlations. 

As noted in the main text, $J_r$ and $J_l$ for BPCC can be established 
independently by using the one-magnon dispersion, and therefore one would wish 
to determine the average spin correlations on the rung and leg bonds directly 
from the experimental dataset. For guidance, by performing ED calculations on 
ladders of coupling ratio $\gamma = 0.39$ and up to $2 \times 12$ spins using 
the ALPS package \cite{alps}, we determine the average rung spin correlations, 
$\langle \mathbf{S}_i^m \cdot \mathbf{S}_i^n \rangle = - 0.6740$, and leg 
correlations, $\langle \mathbf{S}_i^m \cdot \mathbf{S}_{i+1}^m \rangle = \langle 
\mathbf{S}_i^n \cdot \mathbf{S}_{i+1}^n \rangle = - 0.1816$. However, because of 
an unknown intensity scale factor, only the ratio of these correlations can be 
extracted from the measured integrated intensities. From the reduced dataset, 
we deduce the ratio $\langle \mathbf{S}_i^m \cdot \mathbf{S}_{i+1}^m \rangle/ 
\langle \mathbf{S}_i^m \cdot \mathbf{S}_i^n \rangle = 0.22(1)$, which compares
favorably with the ratios 0.2586, given within the bond-operator formalism 
\cite{rnr}, and 0.2694 obtained from ED. The structure factors obtained in 
the SMA by using these average bond correlation values are entirely consistent 
with the measured intensities, as shown by the black, dashed lines in 
Figs.~1(c), 1(g), and 1(i).

\subsubsection{Density-Matrix Renormalization Group}

DMRG calculations of the dynamical structure factor of an ideal spin ladder 
were performed in both the $q_y = 0$ and $\pi$ sectors using the optimized 
exchange interaction parameters $J_r$ = 0.295 meV and $J_l$ = 0.115 meV. The
calculations used the time-dependent DMRG method \cite{Vid04,WF04,DV04,Sch11} 
to determine the spatial spin correlation functions at zero temperature in 
real-time for a ladder of 200 rungs. They employed a fixed maximal number of 
a few hundred states and a maximal simulation time of 200 $\hbar/J_r$, with a 
time-step typically taken as $\delta t = 0.2 \hbar /J_r$. These choices 
ensured that the remaining errors were sufficiently small as to have no 
discernible effect on the primary spectral structures. The dynamical spin 
correlations in momentum space were obtained by Fourier transformation of 
the spin correlation functions. More details of the calculational procedure 
are described in Refs.~\cite{Bou11,rbu}. 

The spectra obtained in this way were convolved with a Gaussian function 
representing the instrumental resolution, combined using Eq.~(\ref{eq}), 
weighted by the magnetic form factor of Cu$^{2+}$, and added to a constant 
background value extracted from the experimental dataset. By this process we 
formed a DMRG dataset for $S({\bf Q},\omega)$ completely equivalent to the 
experimental one. This DMRG dataset was cut, integrated in ${\bf Q}$, and 
integrated in $\omega$ using precisely the same techniques as described 
above. The results are displayed both as the color contours in Figs.~1(b) 
and 2(b) of the main text [for $S({\bf Q},\omega)$] and as the solid lines 
in Figs.~1(c), 1(g), 1(i), and 2(c) [for $I({\bf Q})$]. 

\subsubsection{High-Order Series Expansions}

The systematic extension of the strong-coupling approximation is the high-order 
series-expansion method, which has been applied previously for the calculation 
of bound states and dynamical correlation functions in quantum spin ladders 
\cite{Zhe01,Kne01}. Here we have used perturbative continuous unitary 
transformations (pCUTs) \cite{Kne00,Kne01} to map the spin ladder order 
by order in $\gamma$ to an effective Hamiltonian, $\mathcal{H}_{\rm eff}$, 
which conserves the number of elementary $S = 1$ excitations. Physically, 
the $\gamma = 0$ ground state is the product state of rung singlets and the 
elementary excitation is a rung triplet; at finite $\gamma$, these triplets 
become dressed by their mutual interactions to yield the elementary magnon 
excitations.

Because $\mathcal{H}_{\rm eff}$ conserves the magnon number, each (interacting) 
few-magnon problem can be addressed separately. Here we have determined the 
one-magnon hopping amplitudes up to order 11 in $\gamma$, the two-magnon 
interaction amplitudes with $S_{\rm tot} = 1$ up to order 10, and the matrix 
elements in the one- and two-magnon channels relevant for $S({\bf Q},\omega)$ 
also to 10th order. All series are essentially converged for $\gamma = 0.39$ 
and no extrapolations are required. The one-magnon sector is readily 
diagonalized by Fourier transformation to give the one-magnon dispersion 
and spectral weight directly. 

\begin{figure}[t]
\includegraphics[width=\linewidth]{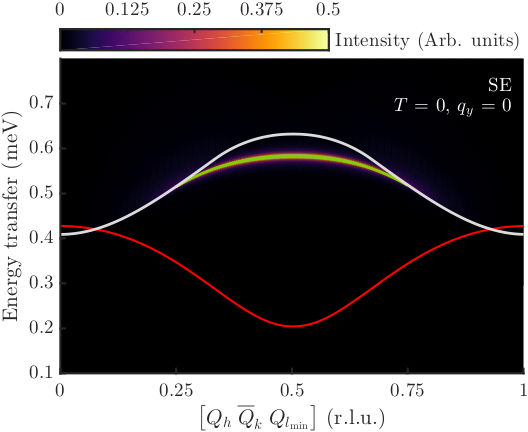}
\caption{\label{hosebs} Dynamical structure factor, $S({\bf Q}, \omega)$, in 
the symmetric ($q_y = 0$) sector, calculated by series-expansion (SE) methods 
up to 10th order, showing the triplet component of the two-magnon bound state 
and the absence of visible intensity from the two-magnon continuum. The red 
line marks the one-magnon dispersion and the white line the lower boundary 
of the two-magnon continuum.}
\end{figure}

In the two-magnon channel, it is necessary to solve a two-body problem to 
obtain $S({\bf Q},\omega)$ \cite{Kne04}, which we execute by Lanczos 
tridiagonalization of the two-magnon $\mathcal{H}_{\rm eff}$ for fixed total 
momentum ${\bf Q}$. The only remaining degree of freedom is the separation, 
$d$, of the two magnons, and by allowing a maximal $d = 4000$ all finite-size 
effects are essentially eliminated from the calculation. From the Lanczos 
tridiagonalization, we extract \cite{Kne04} the complete two-magnon 
contribution to $S({\bf Q},\omega)$, the dispersion and the intensity of the 
two-magnon bound state as a function of ${\bf Q}$, and the corresponding 
intensity of the two-magnon continuum. We note that the two-magnon structure 
factor, $I({\bf Q})$, corresponds to the sum of bound-state and continuum 
contributions and is obtained directly from the matrix elements.

As for our DMRG calculations, to obtain a dataset comparable with experiment 
we scale and smooth our series-expansion data for $S({\bf Q},\omega)$ and 
integrate them in the same manner. Focusing on the results for the two-magnon 
triplet bound state, in Fig.~\ref{hosebs} we show $S({\bf Q},\omega)$ 
calculated in the $q_y = 0$ sector and in Fig.~2(c) of the main text the 
corresponding $I({\bf Q})$ for comparison with the results from experimental 
measurements and DMRG calculations. 

\subsubsection{Two-Magnon Bound States}

For an accurate description of the two-magnon bound state we employ the 
strong-coupling series expansion \cite{Zhe01,Kne01}. For illustration, the 
energy of the triplet component of the bound state is given to third order by  
\begin{eqnarray}
\label{eqn:tripletboundstate}
\frac{\omega_{B}(q)}{J_r} & = & 2 \! - \! \frac{3}{2} \gamma \! + \! 
\frac{11}{8} \gamma^2 \! + \! \frac{17}{16} \gamma^3 \! - \left( \! 
\gamma \! + \! \frac{1}{4} \gamma^2 - \frac{9}{16} \gamma^3 \! \right) 
\cos q \nonumber \\ & & - \frac12 (\gamma^2 + \gamma^3) \cos 2 q
 - \frac{5}{16} \gamma^3 \cos 3 q.
\end{eqnarray}
In the same approximation, the lower edge of the two-magnon continuum is 
located at the energy
\begin{eqnarray}
\label{e2tce}
\frac{\omega_{2m}(q)}{J_r} & = & 2 \! + \! \frac{3}{2} \gamma^2 \! + \! 
\frac{3}{4} \gamma^3 \! - 2 \left( \! \gamma \! - \! \frac{1}{4} \gamma^3
\! \right) \! \cos {\textstyle \frac12} q \nonumber \\ & & - \frac12 
(\gamma^2 + \gamma^3) \cos q - \frac{1}{4} \gamma^3 \cos {\textstyle 
\frac{3}{2}} q.	
\end{eqnarray}
One observes that the triplet bound states are cut off by the presence of the 
continuum, but exist over a range of wave vectors $q_c \le q \le 2\pi - q_c$ 
around the edges of the Brillouin zone, where $q_c = 2 \pi Q_c$ is given in 
Eq.~(1) of the main text. For ladders with stronger rung-to-leg coupling ratio 
(weaker $\gamma$), $q_c \simeq 2\pi/3$ and the bound state is restricted to 
the outer 1/3 of the 1D Brillouin zone. For larger $\gamma$ values, the range 
over which the bound-state mode exists (1) becomes slightly larger \cite{Zhe01} 
and so does its separation in energy from the continuum edge. However, the 
primary factor affecting its visibility is the intensity of the two-magnon 
continuum, which rises significantly more strongly than the energetic 
separation. The very weak two-magnon continuum [Figs.~\ref{hosebs} and 2(c)] 
is therefore the reason why the bound state is so much more clearly visible 
in BPCC than in any other systems studied previously. 

Quantitatively, as discussed in the main text we find that, even for 
a coupling ratio as apparently modest as $\gamma \simeq 0.4$, a 10th-order
expansion is required for an accurate account of the relative positions of 
the bound state and the continuum edge for a pair of magnons. In our analysis 
of the scattering intensities in Fig.~2(c), we compute the structure factor 
by integrating over an energy window up to the two-magnon continuum 
boundary and attribute these contributions to the bound state. The energy 
window above the continuum edge is attributed to two-magnon scattering 
states. It is clear that applying the same process to our $S({\bf Q},
\omega)$ data, calculated both by DMRG and by series expansions, gives an 
excellent quantitative account of the measured intensities. 

We remark here that one of the open questions under these circumstances 
concerns the ``termination'' of the bound-state mode where it meets the 
continuum at the cut-off wave vector, $q_c$. As noted in the main text, 
the termination of the one-magnon mode in the two-magnon continuum, as 
discussed in Ref.~\cite{Sto06}, is precluded in BPCC by their opposite 
parities. However, both bound and scattering states of two magnons appear 
in the same parity sector ($q_y = 0$), and thus there is no parity protection 
for the bound state in this situation. To understand whether the bound state 
ceases to exist as a well-defined excitation of the system at $q_c$, before 
it, or persists in some form into the two-magnon continuum, it is necessary 
to analyze the line width of this spectral feature; a structure-factor 
analysis of the type performed in Fig.~2(c) does not address the origin 
of the intensity contributions. Unfortunately, the data quality of our 
experimental measurements is not sufficient for a quantitative discussion 
of this issue. Numerically, both our DMRG and series-expansion calculations 
may be used to investigate the nature of this termination, which, however, 
lies beyond the scope of the present analysis. 

\begin{figure}[t]
\includegraphics[width=\linewidth]{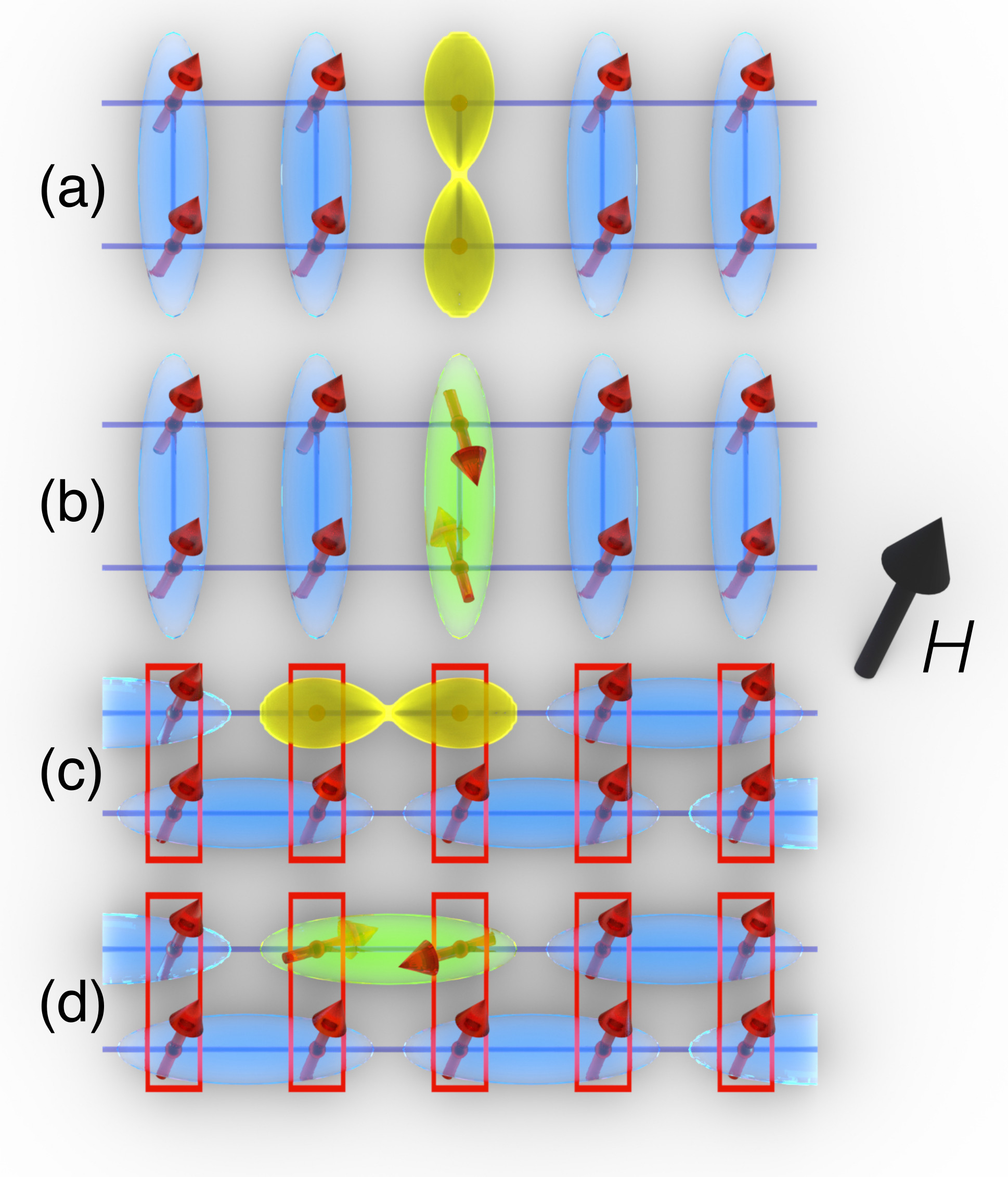}
\caption{\label{fshfp} Schematic representation of the rung states 
and their symmetries in fully field-polarized spin-ladder and Haldane-chain 
systems. (a) Polarized two-leg ladder with one rung-singlet excitation. (b) 
Polarized ladder with one rung-triplet ($t_0$) excitation. Polarized Haldane 
chain in the AKLT representation containing (c) one bond-singlet excitation 
and (d) one $t_0$ bond-triplet excitation.}
\end{figure}

\subsubsection{Field-Polarized Phase} 

The most straightforward description of the fully polarized (FP) phase is 
obtained by using the bond-operator description. For magnetic fields beyond 
the saturation field, $h_s = g \mu_{\rm B} H_s$, there are one singlet and two 
triplet excitations, while the third triplet, $t_+$, is the ground state 
and therefore is treated as fully condensed, i.e. $\langle t_+ \rangle
 = {\overline t}$ is a constant. From the Hamiltonian,
\begin{eqnarray}
H & = & \sum_i \left[ (- {\textstyle \frac{3}{4}} J_r - \mu) s_i^\dag s_i + 
({\textstyle \frac{1}{4}} J_r - \mu + h) t_{+i}^\dag t_{+i} \right. \\
& & + \left. ({\textstyle \frac{1}{4}} J_r - \mu) t_{0i}^\dag t_{0i} + 
({\textstyle \frac{1}{4}} J_r - \mu - h) {\overline t}^2 + \mu + 
{\textstyle \frac{1}{2}} J_l {\overline t^4} \right] \nonumber \\ 
& & + {\textstyle \frac{1}{2}} J_l \sum_i \left[ {\overline t}^2 (s_i^\dag s_{i+1} 
 + t_{0i}^\dag t_{0i+1} + {\rm H.c.}) - 2 {\overline t}^2 t_{-i}^\dag t_{-i} \right], 
\nonumber
\label{ehfp}
\end{eqnarray}
one may extract the mean-field equations  
\begin{eqnarray}
\frac{\partial H}{\partial \mu} & = & 0 \, = \, 1 - {\overline t}^2, \\ 
\frac{\partial H}{\partial {\overline t}^2} & = & 0 \, = \, {\textstyle 
\frac{1}{4}} J_r - \mu - h + J_l {\overline t}^2,
\label{ehmf}
\end{eqnarray}
and hence conclude that ${\overline t} = 1$ exactly, which is the bond-operator 
expression of the fact that quantum fluctuations are completely suppressed in 
the saturated regime. This has the important consequence that the elementary 
excitations are truly non-interacting and hence the exchange parameters 
deduced from their dispersion may be used as a benchmark for cases where 
interaction effects are unknown \cite{Col02}. 
     
Further, the chemical potential, $\mu = {\textstyle \frac{1}{4}} J_r + J_l
 - h$, is governed only by the applied field, $h$ (contributions from the 
sinusoidal excitation bands sum to zero). Thus the dispersion relations are 
\begin{eqnarray}
\!\! \omega_s(q) & = & - {\textstyle \frac{3}{4}} J_r - \mu + J_l 
{\overline t}^2 \! \cos k = h - J_r - J_l (1 \! - \! \cos q), \nonumber \\ 
\!\! \omega_0(q) & = & {\textstyle \frac{1}{4}} J_r - \mu + J_l 
{\overline t}^2 \! \cos k = h - J_l (1 - \cos q), \\ \!\! \omega_-(q) 
& = & {\textstyle \frac{1}{4}} J_r - \mu + h - J_l {\overline t}^2 = 2 h - 
2 J_l. \nonumber
\label{edr}
\end{eqnarray}
The rung-singlet excitation, $\omega_s(q)$, is represented schematically 
in real space in Fig.~\ref{fshfp}(a) and the lowest rung-triplet mode, 
$\omega_0(q)$, in Fig.~\ref{fshfp}(b). The upper mode, $\omega_-$, is 
completely non-dispersive (the band narrows to zero as $h$ increases to 
$h_s$). The lower mode satisfies $\omega_s(q) = 0$ at the band minimum at 
the saturation field, defining $h_s = J_r + 2 J_l$, as expected from simple 
considerations of saturating all the bonds at a single site (no quantum 
fluctuations and thus no correlation effects). These expressions allow a 
very accurate fit of the exchange parameters in the FP regime, as reported 
in the main text, with no requirement for any extra terms in the magnetic 
Hamiltonian beyond those of the pure Heisenberg spin ladder. The small 
deviations of the high-field $J_r$ and $J_l$ parameters from their zero-field 
values are magnetostriction effects, which as in BPCB are weak \cite{Thi09a}. 
Figures \ref{fshfp}(c) and \ref{fshfp}(d) illustrate the close similarity of 
the FP-phase excitations of the Haldane chain to those of the two-leg ladder.